\documentclass[%
 reprint,
 amsmath,amssymb,
 aps,
]{revtex4-2}

\usepackage{graphicx}
\usepackage{dcolumn}
\usepackage{bm}
\usepackage{amsmath}
\usepackage{color} 
\usepackage{soul}

\begin{document}
\raggedbottom

\title{High Phase-Space Density of Laser-Cooled Molecules in an Optical Lattice}%


\author{Yewei Wu}
\email{yewei.wu@colorado.edu}

\author{Justin J. Burau}

\author{Kameron Mehling}

\author{Jun Ye}

\author{Shiqian Ding}
\altaffiliation{Present address: Ludwig-Maximilians-Universität München, Am Coulombwall 1, 85748 Garching, Germany}

\affiliation{
JILA, National Institute of Standards and Technology and the University of Colorado, Boulder, Colorado 80309-0440 \\ 
Department of Physics, University of Colorado, Boulder, Colorado 80309-0390, USA
}

\date{\today}

\begin{abstract}
We report laser cooling and trapping of yttrium monoxide (YO) molecules in an optical lattice. We show that gray molasses cooling remains exceptionally efficient for YO molecules inside the lattice with a molecule temperature as low as 6.1(6) $\mu$K. This approach has produced a trapped sample of 1200 molecules, with a peak spatial density of $\sim1.2\times10^{10}$ cm$^{-3}$, and a peak phase-space density of $\sim3.1\times10^{-6}$. By adiabatically ramping down the lattice depth, we cool the molecules further to 1.0(2) $\mu$K, twenty times colder than previously reported for laser-cooled molecules in a trap. 
\end{abstract} 

\maketitle

Ultracold molecules~\cite{carr2009cold,bohn2017cold} provide opportunities for many important applications ranging from quantum simulation~\cite{karra2016prospects,ni2018dipolar,blackmore2018ultracold} to precision tests of fundamental physics~\cite{safronova2018search,demille2017probing}. Applications in quantum information processing~\cite{demille2002quantum,yelin2006schemes,andre2006coherent}, many-body quantum system simulation~\cite{micheli2006toolbox}, and quantum-state-controlled chemistry~\cite{krems2008cold,balakrishnan2016perspective} benefit directly from a high phase-space density (PSD). Bialkali polar molecules associated from ultracold atoms have been brought to quantum degeneracy~\cite{de2019degenerate} and have enabled the study of dipolar evaporation~\cite{valtolina2020dipolar} and collisional shielding~\cite{matsuda2020resonant}. Besides assembling bialkalis, there has been rapid progress in direct laser cooling and trapping of molecules over the past decade. Magneto-optical traps of molecules have been demonstrated for SrF~\cite{barry2014magneto}, CaF~\cite{truppe2017molecules,anderegg2017radio}, YO~\cite{collopy20183d} in 3D, and CaOH~\cite{baum20201d} in 1D. Sisyphus-type gray molasses cooling (GMC) together with velocity-selective coherent population trapping was applied to cool molecules to 4$-$10 $\mu$K in free space~\cite{cheuk2018lambda,caldwell2019deep,ding2020sub}. These molecules have since been loaded into magnetic traps~\cite{mccarron2018magnetic,jurgilas2021marcollisions,jurgilas2021aprcollisions}, optical dipole traps~\cite{cheuk2018lambda} and optical tweezer arrays~\cite{anderegg2019optical,cheuk2020observation,anderegg2021observation}.

Once molecules are confined in a conservative trap, a key question is how to cool them further for enhanced PSD. 
An optical trap is often small in terms of both the trapping volume and  depth, and is thus suitable for loading an ultracold and highly compressed molecular sample. Optical traps can also offer trapping potentials that are nearly state independent, which is critical for cooling molecules with complex internal structures inside the trap. Thus, an important issue to explore is the differential a.c. Stark shifts between different trappable states to remove potential impediments to further cooling. 

In this Letter, we report trapping of laser-cooled yttrium monoxide (YO) molecules in a 1D optical lattice with a high phase-space density. Despite the complex molecular structure, gray molasses cooling works efficiently for YO inside the lattice.
About 1200 molecules are loaded into the lattice with a temperature of 6 $\mu$K and a lifetime of nearly 1 s. 
This enables us to produce optically trapped molecular samples that are 200 times higher in spatial density and 100 times higher in phase-space density than our previous result in free space~\cite{ding2020sub}.
The molecules are cooled further down to 1 $\mu$K, the lowest temperature achieved for laser-cooled molecules so far, by adiabatically ramping the lattice depth to a smaller value.  

\begin{figure*}[htpb]
\centering
\includegraphics[width=1.9 \columnwidth]{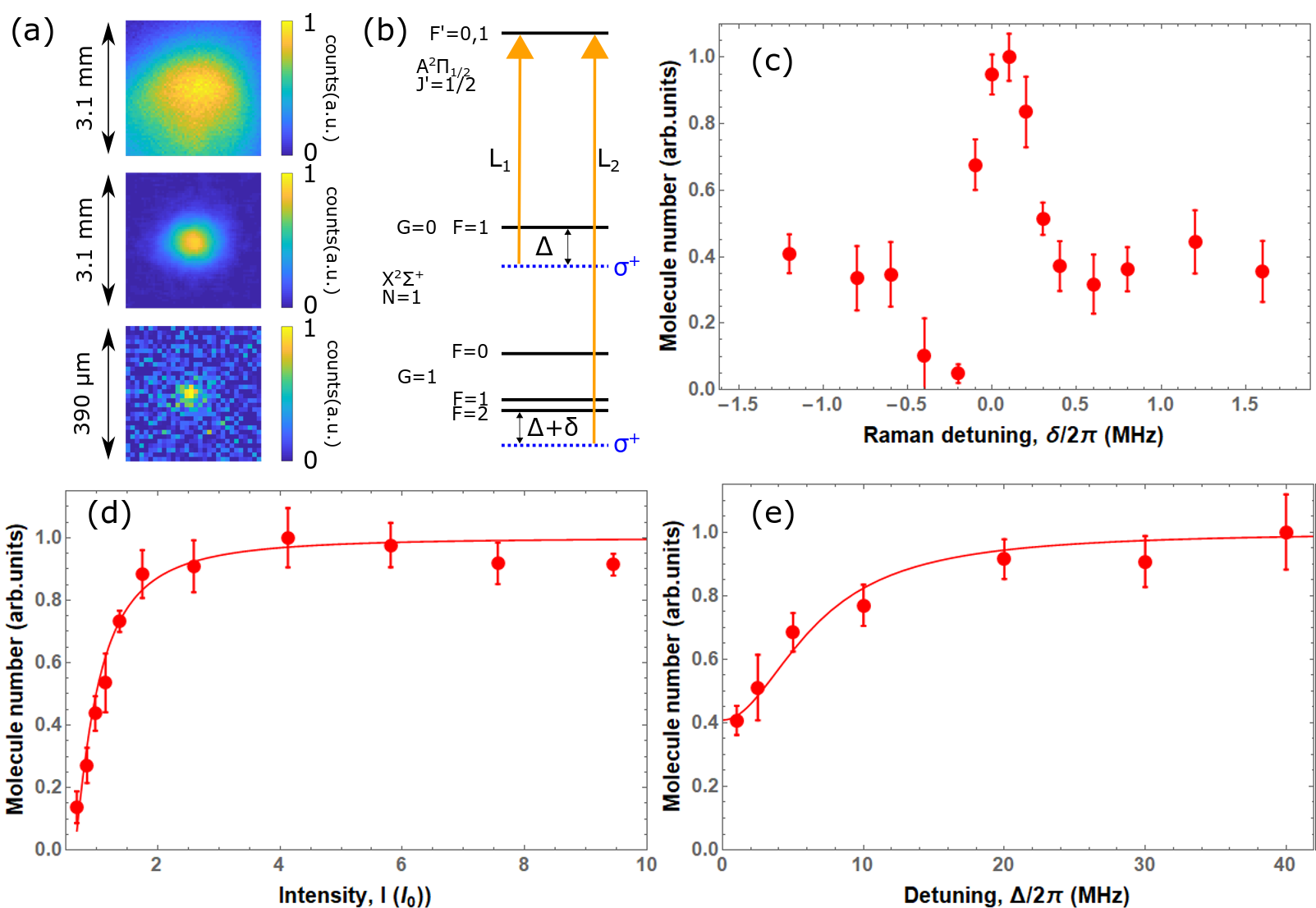}
\caption{\label{fig:figureone}
1D optical lattice loading aided by GMC.  (a) YO cloud at different stages. From top to bottom are dual-frequency MOT, compressed MOT assisted by GMC, and lattice. (b) GMC configuration of YO. The cooling beams consist of two frequency components addressing $G=0$, $F=1$ and $G=1$, $F=2$ manifolds. The single-photon and two-photon detuning are denoted by $\Delta$ and $\delta$, respectively. (c) Number of loaded molecules versus two-photon detuning $\delta$ with $I$=14$I_0$ and $\Delta$=8.3$\Gamma$, where $I_0$=2.7 mW/cm$^2$ is the estimated saturation intensity, and $\Gamma=2\pi \times 4.8$ MHz is the natural linewidth of $A^2\Pi_{1/2}$. (d) Number of loaded molecules versus GMC intensity $I$ with $\Delta$=8.3$\Gamma$ and $\delta$=2$\pi\times$70 kHz. (e) Number of loaded molecules versus $\Delta$ with $I=5.5I_{0}$ and $\delta$=2$\pi\times$70 kHz. 
The solid lines in (d) and (e) are fits to $N=N_{0}-k\times\rho_{ee}$, where $N_0$ is a constant, $k$ is a scale factor, and $\rho_{ee}$ is the excited state $A^{2}\Pi_{1/2}$ population (as a function of $I$ in (d) and $\Delta$ in (e)) calculated with a three-level model (see Ref.~\cite{janik1985doppler}) using the corresponding experimental parameters.
The error bars in the plots correspond to $1\sigma$ statistical uncertainty. 
}
\end{figure*}

The level structure of YO has been detailed in our previous works~\cite{hummon2013PRL,yeo2015PRL,collopy20183d,ding2020sub,LanCheng}.
Laser cooling scheme employs the $X^2\Sigma^+(N=1) \xrightarrow{} A^2\Pi_{1/2} (J'=1/2)$
transition, where $N$ is the rotation quantum number, and $J'$ denotes the total angular momentum
excluding nuclear spin. In both states, the single valence electron is mostly localized at the yttrium atom, which forms an optical cycling center~\cite{CyclingCenter} that can scatter enough photons for slowing and cooling molecules with two vibrational repumpers. 
In $X^2\Sigma^+$, the strong hyperfine (Fermi contact) interaction couples the valence electron spin $\mathbf{S}$ and the yttrium nuclear spin $\mathbf{I}$, and forms an intermediate $\mathbf{G}=\mathbf{S}+\mathbf{I}$. 
The manifold of hyperfine levels $G=0$, $F=1$, where $F$ denotes the total angular momentum, has a negligible Land\'e g-factor, since the direction of the electron spin is not defined in the lab frame.
This feature enables efficient dual-frequency magneto-optical trapping and exceptionally robust gray molasses cooling against various experimental imperfections.
In particular, GMC still works efficiently in the presence of a large magnetic quadrupole  field of the magneto-optical trap (MOT), which allows us to compress the molecular cloud by combining the spring force of MOT and the low temperature of GMC. 

The experiment starts with a dual-frequency MOT, which captures 5$\times10^4$ molecules with a temperature of 2 mK and a spatial density of $6\times10^5$ cm$^{-3}$, as shown in Fig.~\ref{fig:figureone}(a). We compress the cloud by alternating between MOT and GMC operations to effectively create a MOT at a much lower temperature, which increases the molecule density to $2\times10^7$ cm$^{-3}$.  The optimized MOT parameters and cloud-compression sequence are described in Ref.~\cite{ding2020sub}.
After optimization for spatial compression of the ultracold gas, we load the molecules into a 1D optical lattice. The lattice is formed by interfering two counter-propagating linearly polarized 1064 nm laser beams focused to a Gaussian beam waist of 50 $\mu$m. A nominal laser power of 11 W produces a trap depth of 60 $\mu$K with transverse (longitudinal) trapping frequencies of $0.44$ kHz ($93$ kHz). 

To improve the loading efficiency of the optical lattice, we apply GMC cooling and lattice trapping simultaneously for 50 ms. This duration is optimized empirically to maximize the number of trapped molecules. Two laser beams of equal power L$_{1}$ and L$_{2}$ are used to address the two hyperfine manifolds of $G=0$ and $G=1$ in the ground state, as depicted in Fig.~\ref{fig:figureone}(b), limiting GMC parameters to two-photon detuning $\delta$, intensity $I$ and single-photon detuning $\Delta$. After the lattice is loaded, we switch off the GMC beams for 40 ms so that the untrapped molecules escape from the detection region. We measure the number of molecules trapped inside the lattice with a 0.5 ms laser pulse resonantly addressing all the ground hyperfine states in the $X^2\Sigma^+(N=1) \xrightarrow{} A^2\Pi_{1/2} (J'=1/2)$ transition.
An electron-multiplying CCD camera is aligned along the longitudinal direction of the lattice to collect the fluorescence of molecules. 

The spatially varying lattice trapping potential gives rise to variations of differential a.c. Stark shift between different hyperfine states arising from vector and tensor polarizability effects. These differential Stark shifts could potentially impede efficient cooling inside the lattice by destabilizing the dark states required for GMC. Thus, a systematic study of the lattice loading and trapping performance as a function of different GMC parameters can shed light to these issues. Figure~\ref{fig:figureone}(c) illustrates the dependence of the number of loaded molecules on the two-photon Raman detuning $\delta$. For highest loading efficiency, the optimal value of $\delta$ is shifted by $\delta_{\text{trap}}=$2$\pi\times$70 kHz from the Raman resonance condition in free space, $\delta$=0.
The differential Stark shifts between $G=0$, $F=1$ and $G=1$, $F=2$ inside the lattice are calculated based on their coupling to $A^2\Pi_{1/2}(J'= 1/2)$ and $A^2\Pi_{3/2}(J' = 3/2)$, and are predicted to be tens of kHz using parameters of the lattice beams, consistent with our observation.
When $\delta$ is detuned away from the shifted, optimal Raman resonance condition by less than the two-photon Rabi frequency $\Omega$, the dark states formed within both $G=0$, $F=1$ and $G=1$, $F=2$ manifolds are destabilized by their cross-coupling. It leads to deteriorated cooling~\cite{ding2020sub}, and, as a result, the number of trapped molecules decreases. When $\delta$ is much larger than $\Omega$, molecules in free space are cooled to a temperature as low as that for Raman resonance condition, indicating that dark states formed within Zeeman sublevels of the same hyperfine manifold are already sufficient to provide optimal cooling. However,
the number of trapped molecules inside the lattice does not increase to the peak level for $\delta=\delta_\text{trap}$. It indicates that the coherence between Zeeman sublevels of different hyperfine manifolds plays a more important role inside the lattice than it does in free-space for GMC cooling. It is thus clear that the spatial variation of differential Stark shifts between sublevels within a hyperfine manifold has compromised GMC cooling, making the Raman resonance condition more prominent for cooling inside the lattice. 

We next investigate the dependence of the molecule number on GMC intensity $I$ with  $\delta=\delta_\text{trap}=2\pi\times70$ kHz. As shown in Fig.~\ref{fig:figureone}(d), the number of loaded molecules increases with $I$ and eventually saturates at $\sim2I_{0}$, where $I_0$=2.7 mW/cm$^2$ is the estimated saturation intensity. 
This observation is consistent with Ref. \cite{cheuk2018lambda} for CaF molecules. Near the Raman resonance condition, a large GMC intensity reduces the temperature sensitivity to $\delta$, easing the tension between the optimal cooling in free space and inside the lattice, and thus improves the lattice loading.  We further find that the number of loaded molecules increases with single-photon detuning $\Delta$ and, as shown in Fig.~\ref{fig:figureone}(e), saturates at $\sim2\pi\times20 \:\text{MHz} =4.2\Gamma$, where $\Gamma=2\pi \times 4.8$ MHz is the natural linewidth of $A^2\Pi_{1/2}$.
This effect is attributed to the molecular excitation to $A^2\Pi_{1/2}$ and the temperature decrease with $\Delta$, as demonstrated in Ref.~\cite{ding2020sub}.  With optimized GMC parameters, $I=1.4I_{0}$, $\Delta/2\pi=40$ MHz and $\delta/2\pi=70$ kHz, we load a maximum of $\sim$1200 molecules into the optical lattice with a temperature of 7.0(6) $\mu$K. It corresponds to a PSD of $3.1 \times 10^{-6}$~\cite{PSD}, the highest among the laser-cooled bulk molecular gases. 

\begin{figure}[htbp]
\centering
\includegraphics[width=1 \columnwidth]{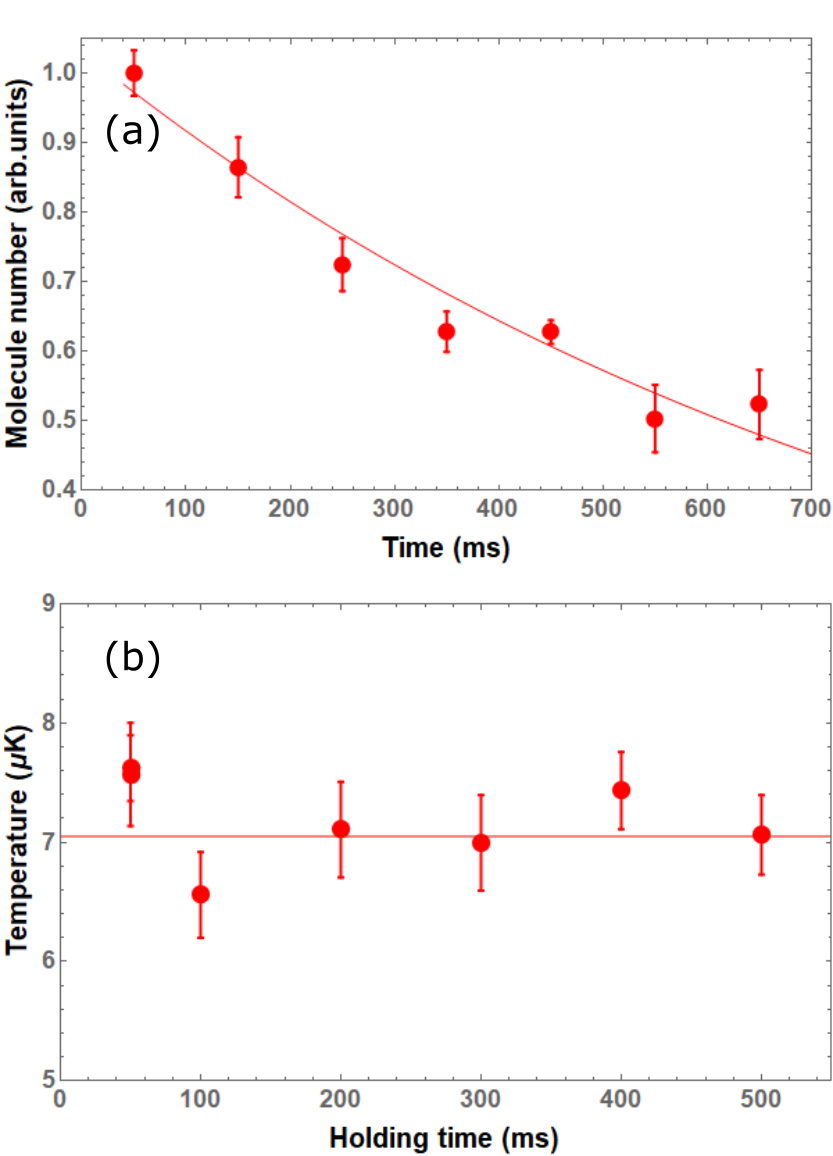}
\caption{
\label{fig:figuretwo}
Characterization of 1D optical lattice. (a) Measurement of lattice lifetime. The solid line is an exponential fit with a $1/e$ decay time of 850(70) ms.
(b) Measurement of molecules heating rate inside the lattice. The solid line is a linear fit with a heating rate of $-0.4(0.7)$ $\mu$K/s.
}
\end{figure}

\begin{figure*}[t]
\centering
\includegraphics[width=1.9 \columnwidth]{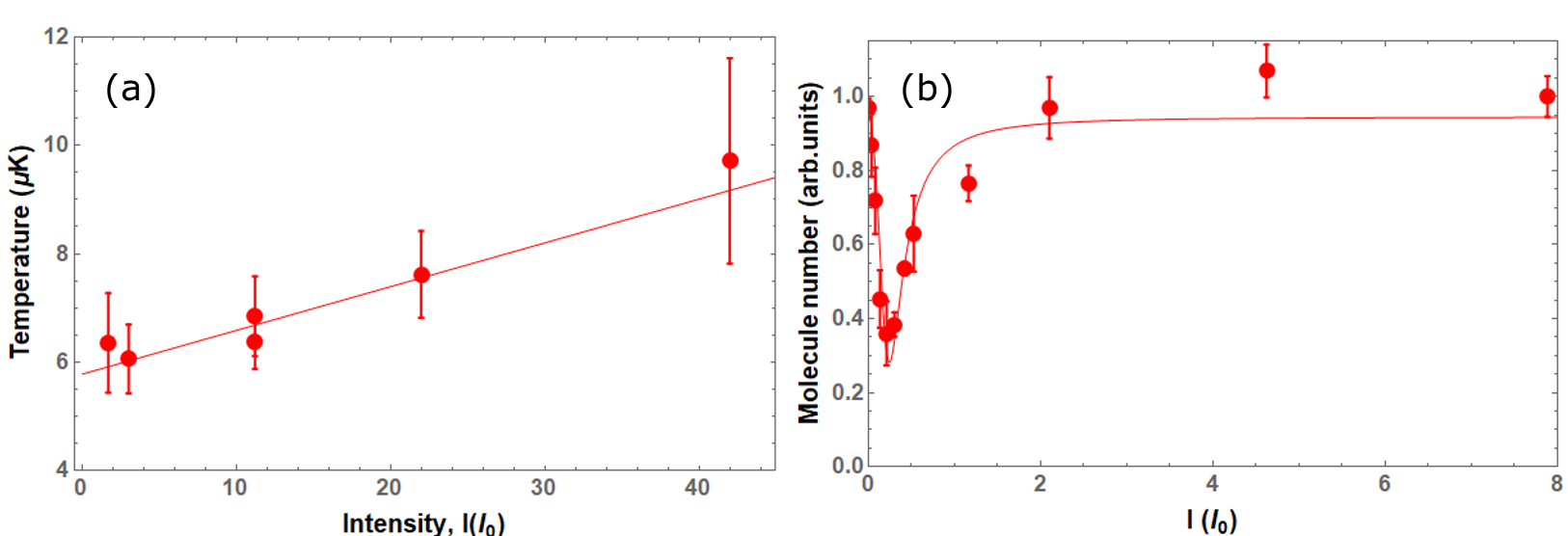}
\caption{\label{fig:figurethree}
(a) Temperature versus GMC intensity $I$ with $\Delta$=8.3$\Gamma$ and $\delta$=2$\pi\times$70 kHz. The solid line is a linear fit. 
(b) Number of remaining molecules versus $I$ with $\Delta$=8.3$\Gamma$ and $\delta$=2$\pi\times$70 kHz. The solid line is a fit to the function $N=N_{0}-k\times\rho_{ee}(I)$, where $k$ is a scale factor, and $\rho_{ee}$ is the excited state $A^{2}\Pi_{1/2}$ population calculated with a three-level model (see Ref.~\cite{janik1985doppler}). Molecules in $A^{2}\Pi_{1/2}$ state could end up in dark states after scattering 1064 nm photons.
}
\end{figure*}

\begin{figure}[htb]
\centering
\includegraphics[width= \columnwidth]{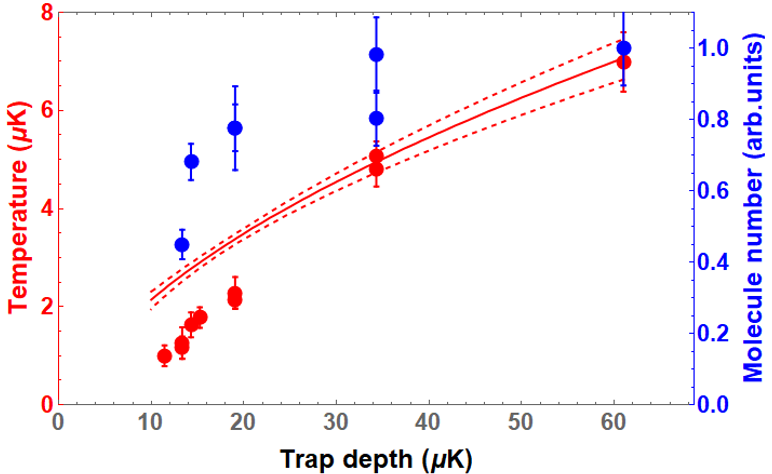}
\caption{\label{fig:figurefour}
Molecule cooling inside the optical lattice by adiabatically ramping down the trap depth. 
The horizontal axis shows the final trap depth after the adiabatic ramping. 
The red and blue dots represent the measured temperatures along the transverse direction and the numbers of the remaining molecules after trap depth ramping, respectively. 
The red line is the calculated temperature based on the remaining molecule number (blue dots) and the assumption of conserved PSD during adiabatic ramping. 
The dashed lines show the 95$\%$ confidence interval. 
}
\end{figure}

To characterize the lattice, we measure the lifetime of trapped molecules and the heating rate induced by the lattice beams. After loading molecules in the lattice, we switch off all other lasers except the lattice beams, and measure the number of remaining molecules after various amounts of time. The results are shown in Fig.~\ref{fig:figuretwo}(a), from which we extract the $1/e$ lattice lifetime to be 850(70) ms.
We attribute the limitation of lifetime to collisions with background gas (vacuum pressure $\sim$3$\times$10$^{-9}$ Torr).
To verify that the molecules are trapped in an optical lattice instead of an optical dipole trap, we rotate the polarization of the retroreflected 1064 nm beam, and observe a dramatic decrease of trap lifetime to $\sim 100$ ms.  The 1064 nm beam is not perfectly horizontal, and the trapping force of an optical dipole trap along the longitudinal direction is not strong enough to hold the molecules against gravity.
To determine the heating rate, we measure how the temperature of molecules evolves over time using the standard ballistic expansion method.
As shown in Fig.~\ref{fig:figuretwo} (b), no noticeable heating is observed within 500 ms.

In our previous study~\cite{ding2020sub}, we showed that temperature of YO gray molasses decreases with light intensity in free space. And this trend holds until light intensity ($\sim$ $1.2I_0$) is too low to hold the molecules against gravity. After the molecules are loaded in the lattice, we ramp down the GMC intensity beyond this limit and keep the GMC beams on for 30 ms to explore the possibility of reaching an even lower temperature. For a large range of $I$, we indeed observe a similar trend of decreasing temperature with smaller $I$ (Fig.~\ref{fig:figurethree}(a)). Strikingly, we also observe a molecular loss near GMC intensity of $I=0.2 I_{0}$, as shown in Fig.~\ref{fig:figurethree}(b). In the presence of lattice light, the Raman resonance condition can not be met for all the molecules. 
When the GMC intensity is sufficiently large that the two-photon Rabi frequency $\Omega$ is much larger than Raman detuning for all the molecules across the lattice beam, the $A^2\Pi_{1/2}$ state population $\rho_{ee}$ is negligibly small. However, for a finite GMC intensity, the differential Stark shift can give rise to a finite $\rho_{ee}$. Numerical simulations based on a three-level model~\cite{janik1985doppler}, with $\delta=2\pi\times 70$ kHz, predict a peak of $\rho_\text{ee}$ near $I=0.2I_{0}$. Once the molecules are excited to $A^2\Pi_{1/2}$, they can be further excited to $D^{2}\Sigma^{+}$ with a 1064 nm photon~\cite{zhang2017high}, which then decays to $X^{2}\Sigma^{+}$. This is a three-photon process and the molecules end up in rotational states with opposite parity compared with the initial $X^{2}\Sigma^{+}$ states in the $X^2\Sigma^+ \xrightarrow{} A^2\Pi_{1/2}$ cycle, and thus they are lost from the cycling transition. The overall shape of $\rho_{ee}$ agrees well with our observation of the intensity-dependent molecular number (see Fig.~\ref{fig:figurethree}(b)).
As a result, the lowest molecular temperature inside the lattice with the presence of GMC is 6.1(6) $\mu$K with $I=3I_0$, $\Delta/2\pi=40$ MHz and $\delta/2\pi=70$ kHz.

We adiabatically reduce the lattice trap depth~\cite{kastberg1995adiabatic,chen1992adiabatic} to further cool the molecules inside the lattice. 
We linearly ramp down the 1064 nm laser power from a trap depth of 60 $\mu$K to various trap depths in 50 ms.
The adiabaticity condition $\lvert\dot{\omega}\rvert/\omega^2\ll1$ is satisfied for both longitudinal and radial trapping directions.
The molecules are held for another 6 ms before the temperatures are measured with the ballistic expansion method. Since the CCD is aligned along the lattice axis, only temperatures along transverse direction are measured. As shown in Fig.~\ref{fig:figurefour}, the temperature decreases as the final trap depth becomes shallower, and we observe a temperature as low as 1.0(2) $\mu$K,  twenty times lower than previously reported for laser-cooled molecules in a conservative trap~\cite{cheuk2018lambda}. Similar observations have been reported in other systems~\cite{kastberg1995adiabatic,chen1992adiabatic}. Assuming that the longitudinal temperature is the same as the transverse one and that PSD is conserved during adiabatic ramping, we calculate molecule temperatures based on the measured numbers of remaining molecules as indicated by the solid curve in Fig.~\ref{fig:figurefour}. The measured temperatures appear to decrease with the trap depth faster than expected.

To conclude, we have demonstrated the first trapping and cooling of molecules in an optical lattice with a phase-space density of 3.1$\times$10$^{-6}$. 
We studied the dependence of lattice loading and trapping on various GMC parameters, and showed that GMC cooling remains efficient inside the lattice with the trapped-molecule temperature as low as 6 $\mu$K. By adiabatically ramping down the lattice depth, we further cooled the molecules to 1 $\mu$K temperature, the lowest for laser-cooled molecules.
With a further increase of molecular number in the lattice we can begin to explore ultracold chemistry and many-body physics with laser-cooled molecules.

We acknowledge funding support from ARO MURI, AFOSR MURI, NIST, and NSF PHY-1734006.

\bibliography{YO_bib}

\end{document}